\documentclass[aps,twocolumn,showpacs,superscriptaddress,amsmath,amssymb]{revtex4}
  
\usepackage{graphicx}  
  
\begin{document}  
  
\title{Theoretical analysis of super--Bloch oscillations}  
\author{K. Kudo}  
\affiliation{Division of Advanced Sciences, Ochadai Academic Production,  
Ochanomizu University, 2-1-1 Ohtsuka, Bunkyo-ku, Tokyo 112-8610, Japan}  
\author{T.S. Monteiro}  
\affiliation{Department of Physics and Astronomy,  
University College London, Gower Street, London WC1E 6BT, United Kingdom}  
  
\date{\today}  
\begin{abstract}  
Several recent  studies have investigated the
 dynamics of cold atoms in optical lattices subject to AC forcing;
 the theoretically predicted renormalization of
 the tunneling amplitudes has been verified experimentally.
 Recent observations include
global motion of the atom cloud, such as giant ``Super--Bloch Oscillations'' (SBOs).
 We show that, in order to understand unexplained features of SBOs,
in addition to the renormalization of the tunneling, a
new and important phase correction must be included.
 For Fermionic systems with strong attractive 
interactions, one may engineer different types of collisions and 
recollisions between bound-pairs and unpaired atoms.  
\end{abstract}  
\maketitle  

\section{Introduction}

Recent experiments on cold atoms in optical lattices subject to time-periodic
 perturbations \cite{Arimondo,Kierig,Eck09,Ferrari,Nagerl,Arimondo2} 
have provided clean 
realizations of a range of different types of AC control of coherent 
matter waves proposed by earlier theoretical studies  
\cite{DL,CDT,Grifoni,Holt,Holt2000,Garreau}. 
Even neglecting the effects of inter-particle interactions, these studies 
identified nontrivial dynamical effects. In particular, for 
an oscillating potential of strength $F$ and frequency $\omega$, 
it was found that the tunneling amplitudes  
$J$ of the driven atoms take an effective, renormalized, value:  
\begin{equation}  
J_{\rm eff}\propto J \mathcal{J}_n\left(\frac{Fd}{\hbar\omega}\right),  
\label{renorm}  
\end{equation}  
where $d$ is the lattice constant and 
$\mathcal{J}_n$ denotes an ordinary Bessel function 
and $n=0,1,2..$. 
Values of $n > 0$ arise if an additional static linear 
(in $x$, the position) field 
$F_0 x$ is applied, satisfying a resonance condition $F_0 d= n\hbar\omega$. 
The inter-site transport is completely suppressed  
 at parameters corresponding to the zeros of the Bessel functions. 
 
The oscillating potential can be implemented  
with ultracold atoms in shaken optical lattices, 
which yield potentials of the form $H_F(t)= F x \sin (\omega t+\phi)$, where 
$\phi$ is a phase which can be controlled. Thus,    
Eq.(\ref{renorm}) was demonstrated and investigated experimentally 
in Refs.~\cite{Arimondo,Kierig}. 
The observed suppression of transport is variously termed Dynamic
Localization (DL) or Coherent Destruction of Tunneling (CDT), 
depending on the transport regime \cite{DL,CDT,Grifoni,Holt2000}. 
They are both effects of the one-particle dynamics;
comparatively less work has been done in the strongly 
interacting regime though, for example, in
~\cite{Holthaus,Creffield,Arimondo2,Malomed} they have investigated 
interactions-driven effects. Recent studies
consider even triangular, shaken lattices \cite{Sengstock}.  
 
 The experiments in Refs.~\cite{Arimondo,Kierig} investigated 
the spreading or local tunneling of an 
atomic wavepacket, without global motion. But an earlier theoretical 
study \cite{Garreau} 
had proposed also the possibility of global transport of the atomic wavepacket
 in the presence of the additional static field and assuming 
$F_0 d \simeq \hbar\omega$ (i.e., $n=1$).
Recent experiments using a static field \cite{Nagerl,Ferrari}
 were able to realize both directed motion as well as large oscillations,
occurring over hundreds of sites, which were termed 
``Super--Bloch Oscillations'' (SBOs) \cite{Nagerl}. These SBOs were 
analyzed in Ref.~\cite{Kolovsky}, including the effects of 
weak interactions 
(mean-field regime). The general assumption was that the group 
velocity in these cases
 $v_g \propto \mathcal{J}_1(F_\omega)$, i.e., 
the dependence of the dynamics on the 
oscillating field $F_\omega\equiv Fd/(\hbar\omega)$, is entirely contained 
in the Bessel function argument. 
 
We show here that, in order to explain the new experiments, 
 a phase correction 
$F_\omega \cos \phi-n(\phi+\frac{\pi}{2})$ 
must---additionally---be considered either
in an effective dispersion relation or in the average group velocities. 
We show that inclusion of the appropriate terms
can account for several unexplained experimental features
in the experiments \cite{Nagerl,Ferrari}. A notable example is a field dependent
shift observed in the phase of SBOs, 
not accounted for by the usual 
analysis. Other features include the different field dependence of 
the speed of directed motion, on
resonance, relative to the amplitude of SBOs.

There is also currently much interest in pairing phenomena, motivated by many 
ground-breaking experiments with ultracold Fermionic atoms in optical 
lattices \cite{Fermi, BP1}.
 We have previously shown \cite{Kudo} that in the case of 
attractive interactions, where
 the atoms can form bound-pairs, the second-order tunneling mechanism implies
 a different renormalization by the field and thus develop a global velocity  
{\em relative} to unpaired atoms. We investigate here for the first time 
the dynamics of pairs in the
SBOs regime. We find the paired/unpaired components can 
be made to  re-collide repeatedly; the novelty here is that the new 
$F_\omega$-dependent 
phase term enables 
control of re-collisions: the two components can either touch and reflect or 
can be forced through each other.  
Experimental demonstration is a matter of combining the techniques
 used to study of dynamics of bound pairs   
\cite{BP1} with the AC driving in 
Refs.~\cite{Arimondo,Kierig,Eck09,Ferrari,Nagerl};
 the re-collisions in two component ultracold gases suggest the possibility of
 other applications in cold chemistry. 

The central objective of the present work is the effect of the phase of
the AC driving acting on an initially undisturbed atomic wavepacket. 
Thus, the phase $\phi$ must be well defined over every cycle,
and the driving field must be switched on initially over a  time that is
 small relative to the driving period. But provided this
constraint is satisfied, the switch-on protocol is immaterial.
The observed effects are never due to a discontinuous ``switch-on'' or an
abrupt jump in amplitude of the potential at $t=0$; in fact
the  strongest experimental effects occur for
$\phi=0,\pi$ where the potential ramps up smoothly from zero at $t=0$ 
(i.e.,  a pure $\sin \omega t$ drive). The authors of Ref.~\cite{Eck07} studied
the effect of adiabatically switching on AC driving over many driving periods
 on the Floquet quasi-energy bands, where the phase is thus of no consequence.
The authors of Ref.~\cite{Creff2} considered slow linear ramping of the driving 
$f(t)=K t \sin (\omega t+\phi)$  by means of a perturbative treatment
valid for high $\omega$. For a ramp period lasting an integer number of periods
$t=nT$, a net ratchet current was obtained.  Other forms
of driving like amplitude modulated lattices that do not show 
dynamic localization can also produce directed motion
  \cite{Alberti}.

In Section II below, we review key aspects of the theory of renormalization
of tunneling for atoms subject to AC forcing.
In Section III we show that the phase corresponds to directed motion and compare
with other systems. In Section IV we analyze Super--Bloch Oscillations 
and obtain an expression 
which we show explains important features of recent experiments.

\section{Renormalization of Tunneling}

As in Refs.~\cite{Eck09,Holt2000}, we
 consider the dynamics in a spatially periodic 
potential, subject to an
 additional time-periodic driving term. The total Hamiltonian is
$H(t) = H_0 +H_F(t)$. 
Here, $H_0$ corresponds to the non-interacting limit of 
a variety of Hamiltonians with nearest-neighbor hopping (Hubbard, Bose-Hubbard, 
magnons in Heisenberg spin chains, etc.). 
It represents any
spatially periodic potential characterized by energy eigenfunctions $\phi_{mk}$,
with band index $m$ and wavenumber $k$, 
thus $H_0 \phi_{mk} = E_m(k)  \phi_{mk}$.
We restrict our one-particle problem, i.e.
\begin{equation}
H_0=-\frac{J}{2}\sum_{j}(c_{j}^\dagger c_{j+1}+ \mathrm{H.c.}),
\end{equation}
where $c_j^\dagger$ and $c_j$ are the creation and annihilation operators 
of a Fermion or Boson,
 to the
lowest band $m=1$; taking $E(k) \equiv E_{m=1}(k)$, the energy dispersion
\begin{equation}
E(k) = -J \cos kd ,
\end{equation}
 so this corresponds to the group velocity 
\begin{equation} 
v_g= \frac{1}{\hbar}\frac{\partial E(k)}{\partial k}
= \frac{Jd}{\hbar} \sin kd. 
\label{vg} 
\end{equation} 
  
Assuming that the external driving is linear in position, we have
$H_F(t) = -f(t) x$  
where $f(t)=F_0+ F \sin (\omega t+\phi)$; 
in general it comprises both a static field
 and a sinusoidally oscillating field with an arbitrary phase $\phi$. 
The result of
 the driving is a time-dependent wavenumber: 
\begin{equation} 
q_k(t)=k +\frac{1}{\hbar} \int_0^t d\tau f(\tau). 
\label{waveq} 
\end{equation} 
The stationary states of the system are its Floquet states, 
the analogues of Bloch waves in a temporally
periodic system. They are given by
\begin{equation}
\psi_k(x,t)=u_k(x,t) \exp\left[-\frac{i}{\hbar} \epsilon(k) t \right] ,
\end{equation} 
where $u_k(x,t)=u_k(x,t+T)$, and the period $T=2\pi/\omega$.
The non-periodic phase term is characterized by
the quasienergy  $\epsilon(k)$. 
The evolution of a wavepacket projected onto its Floquet states 
is fully determined: the quasienergies play a role entirely 
analogous to the energy eigenvalues of a time-independent 
system; the (period-averaged) group velocity $v_g$ of a wavepacket is given 
from their dispersion: 
\begin{equation} 
v_g(k_0)= \left. \frac{1}{\hbar}\frac{\partial \epsilon(k)}{\partial k} \right|_{k_0}, 
\label{vg1} 
\end{equation}
evaluated at the appropriate initial momentum, $k_0$, 
in analogy to Eq.~(\ref{vg}) for the undriven system. 
Below we take $\hbar=1$, which implies $F_\omega=Fd/\omega$. 
 
The presence of the static linear field, in general, destroys the  
band dynamics; however, here we consider the so-called 
resonant driving case, for
 which $F_0 d= n \omega$, 
(where the driving compensates for the energy offset 
between neighboring wells in the
 lattice, restoring the band structure). 
In this case, it can be shown ~\cite{Holt2000} that  
an effective quasienergy dispersion is obtained from the 
energy dispersion by a period-averaging, over one oscillation: 
\begin{equation} 
\epsilon(k) = \frac{1}{T} \int^T_0 E(q_k(t)) dt. 
\label{quasi} 
\end{equation} 

We begin by considering the case $n=0$ : in the first
 experimental studies on Dynamic Localization, $F_0=0$ and thus
the static field was absent. In previous theoretical studies 
(Ref.~\cite{Eck09})
a driving term of form $\sin (\omega t +\phi)$ was considered, but
because of the particular objectives of that work, the effects of the phase
$\phi$ were not retained. In that case, evaluating  the integral in 
Eq.~(\ref{quasi}),
 the  well-known renormalization expression (see also detailed derivation 
in Ref.~\cite{Eck09})
was obtained:
\begin{equation} 
\epsilon(k) =  
-J\mathcal{J}_0(F_\omega) \cos kd.
\label{Renorm} 
\end{equation} 

It can be seen that the tunneling amplitude is multiplied by a Bessel function.
Now the hopping can be completely suppressed at the zeros of the zero-th order Bessel
function. This process was first demonstrated in Ref.~\cite{Arimondo}.
However, for the typical initial wavefunction corresponding to a zero-momentum
ultracold atom cloud, for example, a Gaussian sharply peaked about $k=k_0= 0$,
Eq.~(\ref{vg1}) indicates an average group-velocity $v_g=0$.

\section{Directed Motion}

The situation is quite different if the integral Eq.~(\ref{quasi}) is 
evaluated without disregarding $\phi$. In Ref.~\cite{Kudo} we found that
even for the case $\phi=0$, a momentum shift in the effective dispersion
results. For the case of general $\phi$, the effective
dispersion relation becomes: 
\begin{equation} 
\epsilon(k) =  
-J\mathcal{J}_0(F_\omega) \cos \left[ kd +F_\omega \cos \phi \right],
\label{Renorm1} 
\end{equation} 
where one sees that there is an $F_\omega \cos \phi$  shift, representing
the average momentum over one cycle. This is no longer equal to
zero. The result of this field-dependent shift is to introduce directed
motion, at constant (period-averaged) group velocity \cite{Kudo}.

Atoms in shaken lattices experience a homogeneous (position independent)
force. 
Independently, theoretical \cite{Rid} and experimental \cite{Wal} studies   
considered the role of phase jumps in driven traps.
These have inhomogeneous forces; for a driving field, 
$V(x,t)= V'(x) \sin (\omega t+\phi)$ classically, the phase effects 
a position-dependent momentum shift $\Delta P(x) \propto V'(x)$. The
 position dependence of $V'(x)$ strongly couples the phase to the dynamics.
$\Delta P(x)$ can generate larger/smaller amplitude oscillations in the 
trap and has been proposed as a means to control the kinetic energy
of the atoms oscillating in the trap.

Despite the different dynamics,
for all the above systems, the effect of the phase vanishes for $\phi=\pi/2$.

In Eq.~(\ref{Renorm1}), the $F_\omega \cos \phi$ phase shift coincides with the value of
the lower bound of the integrand in Eq.~(\ref{waveq}) (here evaluated at
 $t=0$). This might lead one to conclude that the shift arises from the
abrupt jump  in the amplitude of the driving potential at $t\simeq 0$. But
this would be a misapprehension;
 the key  physical significance of the shift is that it is the average
momentum over each cycle,  
since $\langle p(t)\rangle_T = F_\omega \cos \phi$ takes the
same form. Below we find that the strongest experimental effects occur for
$\phi=0,\pi$ where the driving grows linearly from zero for  $t\simeq 0$ (a
pure $\sin \omega t$ drive). The phase must be well defined over every cycle,
so  the switch-on time $t_0$ should satisfy the condition $t_0 \ll T$. 
But provided this
constraint is satisfied, the switch-on protocol is immaterial.

We now analyze recently discovered large scale
oscillations for cold atoms in optical lattices. 
Although they are phase-dependent,
we note that they are quite different from the atoms in traps:
the dynamics are independent of initial position (the wavepacket is
delocalized over several wells). 
We show also that the SBOs rely crucially
on a specific quantum resonance (more precisely a slight detuning from it). 
We show below that their amplitude does not depend on $\phi$
(unlike oscillations in traps); only their phase does.

\section{Super-Bloch Oscillations} 

Below we also consider the case of non-zero $F_0$ as  well as slight detuning for which 
$F_0 d= (n+ \delta) \omega$, with $\delta \ll 1$, associated  
with SBOs, for which the above relation still holds. 
    
In order to calculate (period-averaged) group velocities, we first evaluate Eq.~(\ref{waveq}): 
\begin{equation} 
q_k(t)=k + \left[(n+\delta)\omega t- 
F_\omega \cos (\omega t + \phi) + F_\omega \cos \phi \right]/d, 
\label{phaseq} 
\end{equation} 
(assuming $q_k(t=0)=k$) then substitute the result in Eq.~(\ref{quasi}). 
However, for the slight-detuning case $\delta \neq 0$, we assume that 
the time-dependence due to the $\delta \omega t/d$ remains negligible 
over one period $T$. Thus, we take it out of the integral and 
Eq.~(\ref{quasi}) becomes 
(see Appendix~\ref{sec:A} for the detailed derivation): 
\begin{equation} 
\epsilon(k) \simeq  
-J\mathcal{J}_n(F_\omega) \cos \left[ kd + \delta \omega t
+F_\omega \cos \phi  - n(\phi + \frac{\pi}{2})\right].
\label{dispersion} 
\end{equation} 
The above represents an effective dispersion 
relation, but which {\em oscillates slowly in time}  with a period 
$T_{\rm SBO}=2\pi/(\delta\omega) \gg T_B$, where $T_B \propto 1/F_0$ 
is the Bloch period.
They correspond to the SBOs investigated by
Refs.~\cite{Garreau,Nagerl,Ferrari,Kolovsky}. Even at resonance 
$ \delta \omega=0$,
Eq.~(\ref{dispersion}) differs from previous expressions by the phase-shifts 
$F_\omega \cos \phi - n(\phi + \frac{\pi}{2})$. 

Evaluating Eq.~(\ref{dispersion}) for the $n=1$ case, we obtain
\begin{equation}
v_g= \frac{\partial \epsilon}{\partial k}\simeq -Jd \cos \left(kd+F_\omega\cos \phi -\phi + \delta \omega t\right) 
 \mathcal{J}_1\!\!\left(F_\omega\right).
\end{equation}
However, experiments \cite{Nagerl} measure the center-of-mass 
position $x(t)= \int_0^t v_g(t') dt'$,
where the inter-site spacing $d \simeq 0.533$ $\mu$m is included 
if $x(t)$ is obtained in $\mu$m. 
In that case we obtain
\begin{equation}
x(t) \simeq  -\frac{Jd}{\delta \omega}  \mathcal{J}_1(F_\omega) 
 \left[ \sin (K_F+ \delta \omega t)  - \sin K_F \right],
\label{position} 
\end{equation} 
where $K_F=kd+F_\omega \cos \phi  - \phi = F_\omega \cos \phi  - \phi$, 
since in experimental situations of
interest here $k \simeq 0$ initially.

Equation.~(\ref{position}) can largely account for the complex dependence 
of experimental
results on $\phi$. 
For instance, it was noted in Ref.~\cite{Nagerl} 
that the experimental phase of the
SBOs depends on the sign of $\delta$, for $k=0$;
this would not be expected without the $F_\omega\cos \phi -\phi$ shifts,
since $\cos (\delta \omega t)= \cos (-\delta \omega t)$
[see Eq.~(\ref{dispersion})]; it was also noted that
the SBO amplitudes scale as $1/(\delta\omega)$; also, for $\phi=0,\pi$, 
the SBO amplitude
was close to a maximum for $F_\omega=1.52 \simeq \pi/2$.
In contrast, we see that for $\delta=0$ (directed motion) the peak velocity 
occurs wherever $\cos (F_\omega) \mathcal{J}_1(F_\omega)$ is a maximum 
(i.e., at $F_\omega \simeq 1$). 
The 
directed motion is in fact almost zero for $F_\omega= 1.52$, the point
 where the SBOs were near their maximum, but the highest experimental directed
motion was given for  $F_\omega \simeq 1$.

\begin{figure}[tb]  
\includegraphics[height=1.5in]{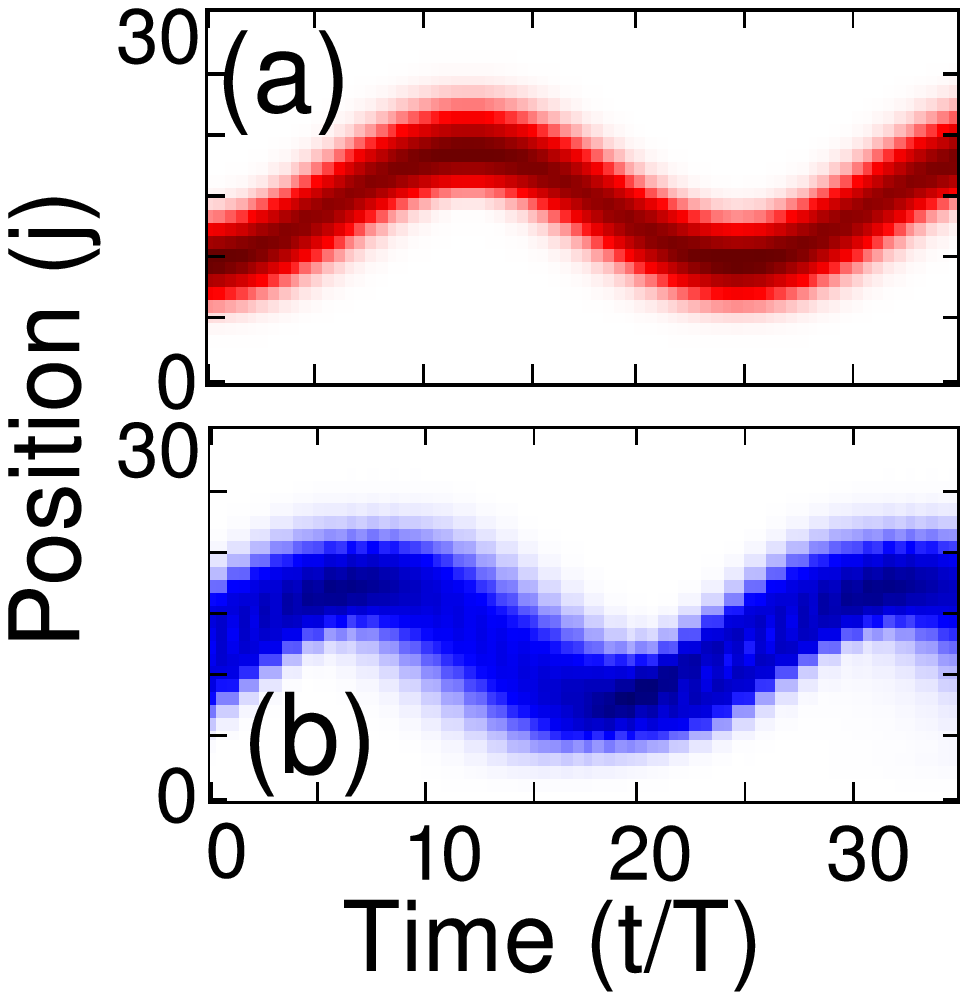}  
\includegraphics[height=1.5in]{Fig1c-d.eps}  
\includegraphics[height=1.5in]{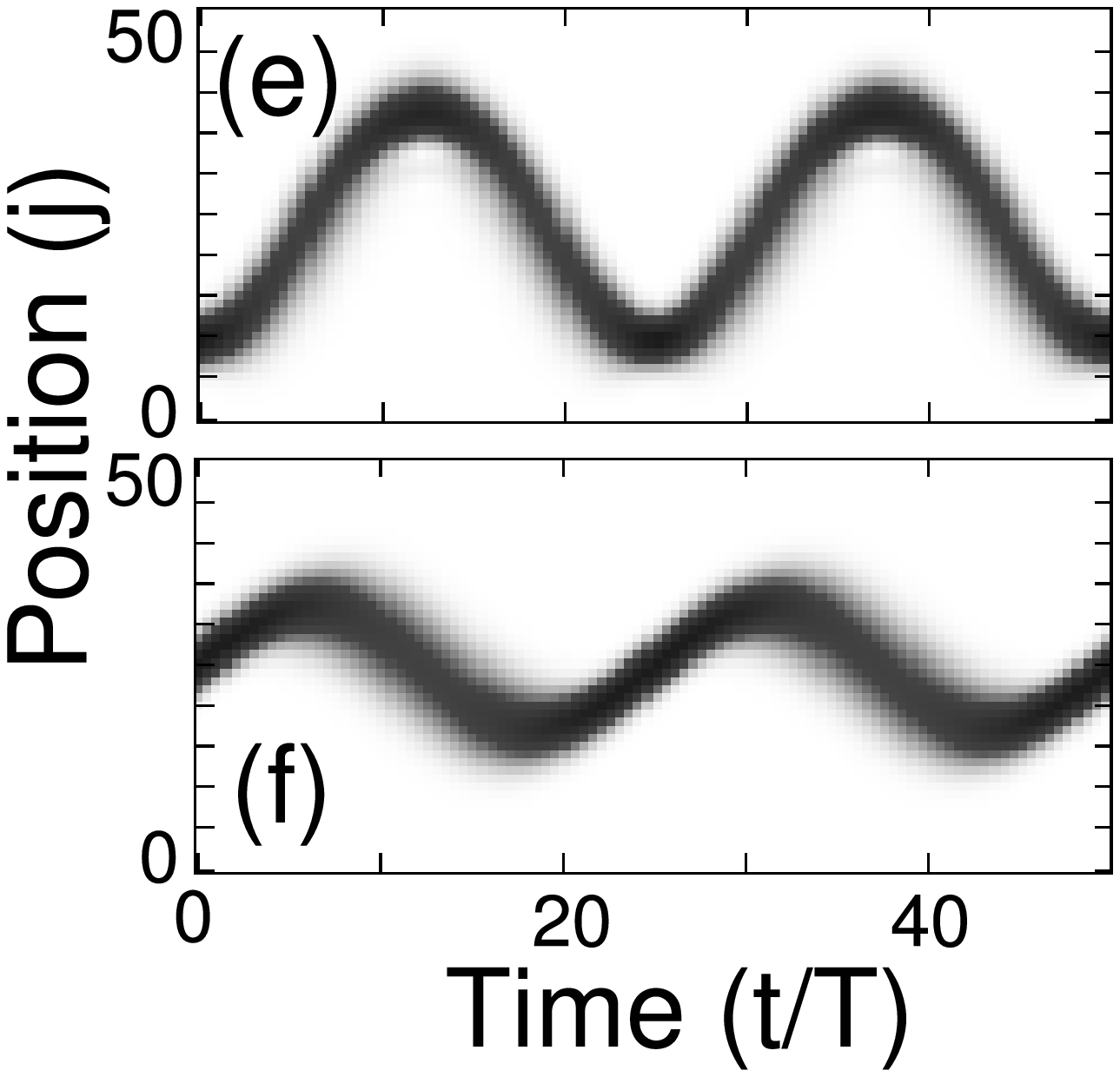}  
\caption{(Color online) {\bf (a)} and {\bf (b)} 
One-particle solutions of the Hubbard
Hamiltonian for $L=30$ lattice sites, showing the field-dependence
of the phase of SBOs.
Upper/lower panel  correspond to  $F_\omega=1.52$ and $0.15$; 
both have $\phi=\pi$.
For illustration purposes, the theoretical amplitudes were equalized 
by choosing $J$-values 
which equalize $J_{\rm eff} \simeq J \mathcal{J}_1(F_\omega)$.
{\bf (c)} and { \bf (b)} Shows that good agreement is 
obtained between experiment
and the Hubbard numerics, as well as the analytical formula: 
Eq.~(\ref{position}) (dashed lines)
reproduces well both the amplitude and phase of SBO  
 experimental data (symbols) of  Ref.~\cite{Nagerl}, using 
$\delta\omega=2\pi/1000$ ($ \equiv -1$ Hz detuning)
 $\phi=\pi$ and $Jd/(\hbar\delta\omega)=90$ $\mu$m. 
This implies $J/\hbar \approx 1.06$ ms$^{-1}$.
 {\bf (e)} and {\bf (f)} show one-particle Hubbard numerics contrasting
  $F_\omega=\pi/2$ (upper panel) with $\pi$ (lower panel),  
parameters used below for collisions with bound-pair states. 
For $F_\omega=\pi$, the wavepacket leaves at high speed
and will return (and recollide with any component) 
at the original position with high speed;
for $F_\omega=\pi/2$ the wavepacket starts with zero speed 
and returns with zero speed.}
\label{Fig1}  
\end{figure}

However, the most interesting experimental feature predicted
by our Eq.~(\ref{position}) is that the SBOs begin with a 
field-dependent phase, 
a feature surprisingly evident even in measurements not looking 
for this behavior. 
Figure~\ref{Fig1} demonstrates this for $F_\omega=0.15,1.52, \pi/2$ and $\pi$, 
in the non-interacting Hubbard model. 
Figures~\ref{Fig1} (c) and (d) show that
Eq.~(\ref{position}) reproduces quite well  
the experimental values of Ref.~\cite{Nagerl} especially for small times. 
The disagreement with experimental data in large times can be attributed
mostly to interactions.
The graph shows
clearly the displacement of the first maximum, seen in the experiment as well
as the order of magnitude variation in amplitude. 
Such a field-dependent shift is also
apparent in the results of Ref.~\cite{Ferrari}. 
Figures~\ref{Fig1} (e) and (f)  contrast $F_\omega= \pi/2$ 
with $F_\omega= \pi$.  
For the former, the atomic wavepacket begins with near-zero speed 
and gradually accelerates,
 while for $\pi$ it leaves with maximal speed (i.e., it acquires the speed on a
timescale $T_{\rm B} \ll T_{\rm SBO}$) and returns to the original 
position with maximal speed.
As we see below, this has interesting physical implications if, 
when the wavepacket returns to
its original position, it re-collides with an immobile component. 

\section{Bound-Pairs}

We now consider the effect of the SBO phases in a regime of attractive 
interactions and Fermionic atoms, such as investigated 
in Refs.~\cite{Fermi,BP1}. In that case, 
we model the dynamics by the many-body Hubbard Hamiltonian:
\begin{equation}
 H_0= -\frac{J}{2} \sum_{j} \sum_{\sigma=\uparrow,\downarrow} 
(c^{\dagger}_{j,\sigma}c_{j+1,\sigma}
+\mbox{\rm H.c.}) + U \sum_j n_{j\uparrow}n_{j\downarrow},
\end{equation}
where $c_{j,\sigma}^\dagger$ and $c_{j,\sigma}$ are the creation and 
annihilation operators of a Fermion and 
$n_{j,\sigma}=c_{j,\sigma}^\dagger c_{j,\sigma}$.  
This supports Bound-Pair (BP) states, lower in energy by $|U|$ relative to 
the unpaired states; as for the unpaired atoms, we consider their motion
 in the lowest band only, with dispersion relation \cite{BPD}: 
 \begin{equation} 
 E(k_1,k_2) = -\frac{J^2}{2U} \cos [(k_1+k_2)d]\equiv -\frac{J^2}{2U} \cos 2\kappa d, 
\label{BP.E_k}
\end{equation} 
where $\kappa$ is the center-of-mass momentum. Similarly, in the   
many-body Hamiltonian, the driving term: 
\begin{equation} 
H_1(t) = -\left[F_0 + F \sin (\omega t + \phi)\right] d \sum_j  j {\hat n}_j,  
\end{equation} 
where $j$ denotes the site index  and ${\hat n}_j$ is the number occupancy 
( $=2$ for a BP; and $=1$ for a site $j$ occupied by either an unpaired spin-up 
or spin-down atom). Thus $H_1^{\rm BP}(t)= 2H_1^{\rm unpaired} (t)$, i.e., the
 magnitude of the driving term is doubled for the BP.

\begin{figure}[tb]  
\includegraphics[width=2.5in]{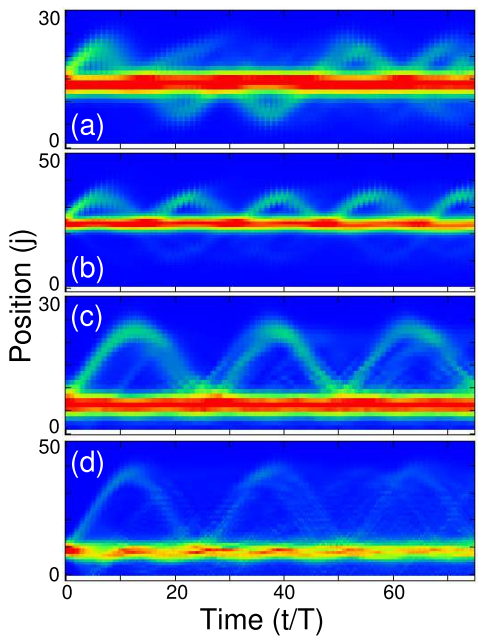}  
\caption{(Color online) Solution of the Hubbard Hamiltonian 
dynamics of a mixture of 
Bound-Pairs and unpaired atoms in the SBOs
regime for $L=30$ or $50$ lattice sites. 
The dynamics are qualitatively similar for $N=2$--$4$ atoms; 
these calculations use $N=3$,
with two spin-up and one spin-down atoms.
Since $J^2d/U \ll 1$, the bound-pair component is almost static 
and appears as a bright horizontal line; 
the free atoms ($1/3$rd of the total probability, initially)
perform SBOs around them.
 {\bf (a)} and {\bf (b)}  correspond to $U=-10$, $F_\omega=\pi$. 
The unpaired atoms separate then recollide with BPs at maximum speed, 
then  attempt to cross the static BP component.
At low energy and large effective $\hbar\propto 1/L$ ({\bf (a)} $J=0.6$, $L=30$), 
there is a mixture of transmission and reflection; 
At higher energy and small $\hbar$ ({\bf (b)} $J=1$, $L=50$), 
complete reflection and a series half SBOs predominate.
 {\bf (c)} and {\bf (d)} are for  $U=-7$, $F_\omega=\pi/2$. 
For $F_\omega=\pi/2$, the wavepacket starts at zero speed 
and returns with zero speed for both $L=30$, $J=0.5$ ({\bf (c)})  
and $L=50$, $J=1$ ({\bf (d)}); there is relatively little transmission.}
\label{Fig2}  
\end{figure}  

Repeating the arguments leading to Eq.~(\ref{dispersion}), we can show that
 for the regime of the experiments ($n=1$ and $\phi=0$) 
\cite{Nagerl,Ferrari}, the BP group velocity is 
\begin{equation} 
v_g^{BP} \simeq -\frac{J^2d}{U} \sin \left(2\kappa d + 2F_\omega +2\delta \omega t\right)\
\mathcal{J}_2\left(2F_\omega\right).  
\label{BP.v_g}
\end{equation} 
Thus, the order of the Bessel, its argument, and the additional phase 
are all doubled;
 so the zeros and maxima of the BP motion occur at different fields 
from the unpaired atoms. 
However, in the presence of the driving, neither the BPs nor unpaired states
 remain eigenstates of the system. For $|U| \gtrsim 2(F_0+ F)$, though, the 
energy gap between BPs and unpaired atoms cannot be closed 
by the fields, and there is   relatively little mixing. 
We tested this assumption for simulations for 
lattices with $L\simeq 30$--$50$ sites and $N=2,3$, and $4$ atoms. 
Numerics are presently unfortunately limited by $N=4$ and $L\simeq 30$; since 
$N=4$ includes the residual BP-BP interactions (and BP's interact like
hard-core bosons), it is expected that the essential physics is included.

Nevertheless, we take a conservative approach and report below only on
dynamics that are qualitatively insensitive to whether  $N=2$ or $N=3,4$.
Thus, we consider the large $U$ regime where BP and unpaired dynamics is 
separated and break-up rates of the BPs are modest: 
for $|U| \lesssim 2(F_0+ F)$,
the pairing rate falls rapidly and can oscillate in time.
A further advantage of large $|U|$ is that the BPs are essentially static,
thus reducing another source of uncertainty.

In Fig.~\ref{Fig2} we show the results of calculations slightly above the
threshold where the field can strongly couple the Bound-Pair states
and free atoms, and there is already a certain degree of interaction in
evidence. 
It is seen that the BP component is essentially immobile, 
while the unpaired atoms perform large-scale oscillations;
the unpaired atoms periodically return
and re-collide with the BP packet. 
However, the $F_\omega=\pi$ trajectories
attempt to ``push-through'' the BPs, leading in general to a degree
of beam-splitting, which depends on the kinetic energy (i.e., $J$) and 
effective $\hbar$ ($\propto 1/L$).
The $F_\omega=\pi/2$ atoms, on the other hand, return with zero velocity
and simply turn around.

Our analysis on BP dynamics is valid only for $|U|\gg\omega\sim J$. 
In this regime, bound pairs remain in the BP band, and thus
Eqs.~(\ref{BP.E_k}) and (\ref{BP.v_g}) are likely to remain valid.
If $\omega \gg |U|$, new types of tunneling might be found.
Such a high-frequency regime warrants further investigation.

\section{Conclusion}

We have shown a phase-correction  in an effective 
dispersion relation is essential for full understanding of
current experiments on transport with AC forcing which showed
Super--Bloch Oscillations.

In addition, we investigate Super--Bloch Oscillations for systems with 
bound pairs.
We show that we can control collisions and re-collisions between unpaired atoms
and bound pairs. 
This has potential implications for studies of AC control of two-component
condensates, including Fermionic systems and molecular condensates,
of relevance in cold chemistry.
     
We acknowledge helpful comments or discussions with A.~Eckardt, 
M.~Holthaus, C.~Weiss  
C.~Creffield, A.~Kolovsky, G.~Ferrari, E.~Arimondo and O.~Morsch. 
We are extremely grateful to 
E.~Haller and C.~N\"agerl for the experimental
data in Fig.~\ref{Fig1}. 
This work is partly supported by KAKENHI(21740289). 

\appendix

\section{\label{sec:A} Effective dispersion relation}

In order to derive Eq.~(\ref{dispersion}) from Eq.~(\ref{phaseq}), 
we assume that the Super--Bloch period represents a completely
different timescale from the much faster Bloch period. Thus, we
 introduce two timescales, a rapid time $t$ for the time-average 
over a period and
a slow time $t' \equiv \delta\omega t$ which describes the slow
dynamics.  Equation~(\ref{dispersion}) is rewritten as
\begin{equation}
 q_k(t)= k + [t' + F_\omega\cos\phi + n\omega t 
- F_\omega\cos(\omega t + \phi)]/d.
\label{a.q_k}
\end{equation}
Since $\delta \ll 1$, so $t'\ll t$ and the change due to $t'$ during 
one period $T$ is negligible. 
Substituting (\ref{a.q_k}) into (\ref{quasi}), we have
\begin{eqnarray}
 \epsilon(k) &=& -\frac{J}{T}\int^T_0 dt
\cos\left[
kd + t' + F_\omega\cos\phi 
\right.
\nonumber\\
&& \left. \hspace{6em}
+ n\omega t - F_\omega\cos(\omega t + \phi)
\right]
\nonumber\\
&=& -\frac{J}{2\pi}\int^{2\pi+\phi}_\phi d\tau
\cos\left[
kd + t' + F_\omega\cos\phi -n\phi
\right.
\nonumber\\
&& \left. \hspace{6em}
+ n\tau - F_\omega\cos\tau
\right]
\nonumber\\
&\simeq& -\frac{J}{2\pi}
\cos(kd + t' + F_\omega\cos\phi -n\phi)
\nonumber\\
&& \quad\quad
\times\int^{2\pi}_0 d\tau \cos[n\tau - F_\omega\cos\tau]
\nonumber\\
&& + \frac{J}{2\pi}
\sin(kd + t' + F_\omega\cos\phi -n\phi)
\nonumber\\
&& \quad\quad
\times\int^{2\pi}_0 d\tau \sin[n\tau - F_\omega\cos\tau],
\end{eqnarray}
where $\tau=\omega t+\phi$.
Evaluating both integrals above, one obtains  Eq.~(\ref{dispersion}):
\begin{eqnarray} 
\epsilon(k) &\simeq&  
-J\mathcal{J}_n(F_\omega) \nonumber \\
&&\times \cos \left[ kd + \delta \omega t
+F_\omega \cos \phi  - n(\phi + \frac{\pi}{2})\right]. 
\end{eqnarray}

The Bessel function in Eq.~(\ref{dispersion})
is obtained from the (non-zero) integrals;
the additional term $n\pi/2$
in the phase arises because the first integral vanishes for $n$ odd, 
while the second
integral vanishes for $n$ even.

\end{document}